\documentclass[10pt]{article}
\usepackage{amsmath,amssymb,amsfonts}
\usepackage{booktabs}
\usepackage{graphicx}
\graphicspath{{pipeline3_images/}}
\usepackage{xcolor}
\usepackage{hyperref}
\usepackage{geometry}
\usepackage{cite}
\usepackage{threeparttable}
\usepackage{algorithm}
\usepackage{algpseudocode}
\newcommand{\repo}{https://github.com/saifibnaezhararko/Accelerating-hybrid-quantum-classical-machine-learning-tasks-on-HPC-platforms}

\title{\textbf{A Hybrid Deep Learning and Variational Quantum Circuit Approach for TREC Question Classification}}

\usepackage{authblk}

\author[1]{SK Saif Ibna Ezhar Arko}
\author[2]{Zijia Zhao}
\author[3]{Maha A. Metawei}

\affil[1]{BRAC University, Dhaka, Bangladesh\\ \protect\href{mailto:saifibnaezhararko1@gmail.com}{\texttt{saifibnaezhararko1@gmail.com}}}
\affil[2]{University of Melbourne, Melbourne, Australia\\ \protect\href{mailto:zhaozj@student.unimelb.edu.au}{\texttt{zhaozj@student.unimelb.edu.au}}}
\affil[3]{High Performance Computing Lab, Electronics Research Institute, Cairo, Egypt\\ \protect\href{mailto:maha\_metawei@eri.sci.eg}{\texttt{maha\_metawei@eri.sci.eg}}}
\date{}

\begin{document}

\maketitle

\begin{abstract}
Hybrid quantum–classical pipelines combine classical feature extraction with variational quantum circuits (VQCs) for text classification. Evaluating these pipelines requires consideration of both classification accuracy and training cost. We present a hybrid convolutional neural network (CNN)–VQC pipeline for binary classification between the description (DESC) and human (HUM) question categories in the TREC dataset.
This study examines the feasibility and computational cost of a hybrid convolutional neural network (CNN) and variational quantum circuit (VQC) for classifying TREC questions into description and human categories. Trainable word embeddings and convolutional layers extract features, which are projected onto four rotation angles and processed by a simulated four-qubit circuit. A classical control replaces the circuit with a linear layer.

The models were evaluated on a balanced subset comprising 174 training and 44 test questions, constructed by pooling and re-splitting the official TREC files. Test questions were restricted to vocabulary present in the training set; consequently, the results are not directly comparable with those obtained using the official test split.
After 20 epochs with a single random seed, the hybrid model using Aer GPU simulation correctly classified 43 of 44 test questions (97.73\%), compared with 42 of 44 (95.45\%) for the classical control. The small test set, single-seed evaluation and differences in training conditions preclude attribution of an accuracy advantage to the quantum layer.
Training required 10,127 seconds for the Aer GPU configuration and 1.3 seconds for the classical control. These timings reflect different execution and training conditions and should not be interpreted as an intrinsic cost of the hybrid architecture. Repeated parameter-shift evaluations likely contributed to the overhead, but profiling is needed to identify the dominant source. The findings demonstrate the feasibility of the approach under simulation while highlighting its configuration-dependent computational costs.

\end{abstract}


\section{Introduction}
\label{sec:intro}
Quantum Natural Language Processing (QNLP) concerns the design and implementation of NLP models intended to run on quantum hardware \cite{lorenz2023qnlp}. The field rests on a formal similarity between the compositional model of meaning and quantum theory, which yields sentence representations with a natural mapping to quantum circuits \cite{lorenz2023qnlp, coecke2020foundations}. The framework common to de facto all QNLP approaches is the Categorical Distributional Compositional (DisCoCat) model, which extends pregroup grammars so that a sentence's grammatical reduction becomes a diagram encoding word interactions \cite{guarasci2022quantum, meichanetzidis2023}. Diagrammatic reasoning then carries the pipeline: categorical quantum mechanics interprets language as quantum processes, and the ZX-calculus translates those diagrams into circuits whose parameterisations become the variables to be learned \cite{coecke2020foundations}. Because the original algorithm presupposed an unrealised quantum random access memory, near-term work instead encodes distributional embeddings as classical ansatz parameters within variational quantum circuits, however the NLP problem scaling emerged as a system bottleneck.
\subsection{Scaling as a Multi-Dimensional Resource Cost}
Scaling is not a single problem: there is more than one way to scale up NLP tasks, and what is distinctive to QNLP is how each dimension surfaces as a resource cost on noisy intermediate scale quantum (NISQ) devices \cite{lorenz2023qnlp}. Sentence length widens the circuit in proportion to the qubits assigned to each pregroup type, but post-selection adds an exponential time cost in the number of qubits measured \cite{lorenz2023qnlp, metawei2023}. Increasing word-meaning dimension makes the problem size grow exponentially and affects classification , while larger vocabularies raise the dimensionality of the parameter space \cite{guarasci2022quantum}. A greater number of sentences instead acts as a multiplicative prefactor on time cost, in principle parallelisable across processors but currently constrained by shot limits, job throughput and processor availability \cite{lorenz2023qnlp}.
Current implementations remain bounded in practice. One self-attention model restricts evaluation to a maximum sentence length of eight words because of simulation and hardware constraints \cite{zheng2026quantum}, and a few-shot NLI benchmark deliberately limits both data and circuit size, training on 100 sentence pairs with one qubit per word in classical simulation \cite{sun2025quantum}. The small number of available qubits constrains the number of sentences, their length and the overall vocabulary span \cite{guarasci2022quantum}, and the scale of applications on actual hardware has not yet matched classical techniques \cite{widdows2024quantum}.\\
\subsection{Pushing the limits with hybrid models}
The practical use of hybrid quantum--classical machine learning models
depends on both predictive performance and training efficiency.
Classical simulation provides a platform for developing and evaluating
these models, but its computational cost can constrain experimentation
\cite{lorenz2023,metawei2023}.
When training uses the parameter-shift rule, gradient computation
requires repeated circuit evaluations, adding to this cost
\cite{mitarai2018,schuld2019}.
These computational demands motivate investigating whether
high-performance computing (HPC), particularly GPU-based quantum
simulation, can improve the efficiency of hybrid model training.
Variational quantum circuits (VQCs) are parameterised circuits whose
gate angles are trained by a classical optimiser.
They form a widely studied approach to machine learning within the
broader context of noisy intermediate-scale quantum computing
\cite{preskill2018,biamonte2017,cerezo2021}.

In quantum natural language processing, one approach constructs
quantum circuits from the grammatical structure of sentences
\cite{coecke2010,meichanetzidis2023,lorenz2023}.
Hybrid quantum--classical variants of this approach have also been
applied to topic-based classification \cite{metawei2023}.
For such grammar-based models, circuit width can grow with sentence
length, increasing the cost of classical simulation.

An alternative approach uses a classical neural network to compress
the input into a compact feature representation, which is then
processed by a fixed-width quantum circuit and a classical output
layer \cite{mari2020}.
Related hybrid architectures have been explored for speech recognition
\cite{yang2021}, sequence modelling \cite{chen2022}, and text
classification \cite{yang2022}.
By separating classical feature extraction from quantum processing,
this approach allows circuit width to remain independent of input
length.

Question classification assigns a question to a category according
to the expected answer type, such as a description or a person,
and is an established component of question answering systems
\cite{voorhees2000,li2002}.
The TREC question classification dataset provides a benchmark for
this task \cite{li2002}.
Convolutional neural networks over word embeddings have demonstrated
strong performance on sentence classification tasks, including TREC
\cite{kim2014,kalchbrenner2014}, motivating their use as the classical
feature extractor in our hybrid model.
We use a balanced binary subset of TREC containing description (DESC)
and human (HUM) questions as a small-scale test case.
To build it, the official TREC training and test files were pooled
and re-split, and test questions were restricted to words that occur
in the training split.
As a result, 42 of the 44 test questions come from the official
training file, and the reported accuracy is not directly comparable
with published results on the official TREC test set.
Section~\ref{sec:sampling} gives the full sampling procedure.

Within this setting, we focus on two evaluation questions.
First, what classification performance does the hybrid CNN--VQC model
achieve relative to a classical CNN baseline?
Second, how does training runtime vary across simulation and gradient
computation configurations, and what benefit does GPU-based simulation
provide within a parameter-shift training workflow?

To address these questions, we construct a hybrid classifier in which
a TextCNN feature extractor maps each question to four rotation angles,
a four-qubit VQC processes these angles, and a classical output layer
produces the classification prediction.
The evaluation uses 174 training and 44 test questions from the
selected subset.
A classical CNN baseline replaces the quantum circuit with a linear
layer followed by a hyperbolic tangent activation.
We train the hybrid model using CPU-based state-vector backpropagation
and GPU-based Qiskit Aer simulation with parameter-shift gradients.
We additionally estimate the cost of CPU-based Aer training from
short timing measurements on a separate workstation.
Differences in gradient computation, batch size and hardware mean
that these comparisons assess the implemented training configurations
rather than isolate the effect of the execution device.

This paper provides a reproducible assessment of the classification
performance and training cost of a fixed-width hybrid model.
In this small-scale evaluation, the observed differences between
execution configurations are substantially greater in training
runtime than in classification accuracy.
Although GPU-based Aer training is faster than the projected CPU-based
Aer training time, the CPU configuration using state-vector
backpropagation is substantially faster than the Aer GPU configuration.
These findings highlight the importance of considering the simulator,
gradient computation method and batching strategy together when
assessing the computational benefit of GPU-based simulation.

The paper is organized as follows.
Section~\ref{sec:related} reviews related work, and
Section~\ref{sec:method} describes the dataset, the sampling criteria,
the model architecture and the training procedure.
Section~\ref{sec:setup} describes the hardware and software, the
training configurations and the evaluation protocol, and
Section~\ref{sec:results} reports the results.
Section~\ref{sec:discussion} discusses the findings,
Section~\ref{sec:limitations} outlines the limitations, and
Section~\ref{sec:conclusion} concludes the paper.

\section{Related Work}
\label{sec:related}
This section surveys previous research across three complementary areas. First, it examines classical approaches to question classification, providing the methodological foundation for the CNN-based feature extractor and a reference point for comparison with previously reported TREC classification results. Second, it reviews quantum and hybrid quantum-classical approaches to NLP classification, encompassing methods that exploit linguistic structures to construct quantum circuits and those that integrate quantum layers with classical text representations. Finally, it discusses the computational complexity associated with gradient evaluation and quantum circuit simulation, establishing the theoretical context for the runtime analysis conducted in this study.

\subsection{Classical Methods for Question Classification}

Early approaches to question classification combined supervised
learning with explicitly designed linguistic features.
Li and Roth \cite{li2002} organised answer types into six broad
categories and fifty fine-grained categories, and developed a
hierarchical classifier using lexical, syntactic and semantic features.
Their experiments on TREC questions demonstrated the contribution
of semantic features to classification performance, particularly
for fine-grained answer types.
However, the hierarchical structure did not consistently improve
performance over a flat classifier.
Zhang and Lee \cite{zhang2003} investigated support vector machines
for TREC question classification and introduced a tree kernel to
incorporate syntactic structure.
Their approach illustrates how structural information can supplement
surface text features, while requiring syntactic parsing of the input.

Later neural approaches learned text representations from word
embeddings, reducing reliance on manually designed features.
Kim \cite{kim2014} evaluated a single-layer CNN across several
sentence classification benchmarks, including TREC.
The CNN-non-static variant, which fine-tuned pretrained word
embeddings, achieved 93.6\% accuracy on TREC.
Kalchbrenner et al. \cite{kalchbrenner2014} reported 93.0\% accuracy
on the same task using a dynamic convolutional network.
These results establish CNNs as relevant classical models for
question classification.
Zhang and Wallace \cite{zhang2017} subsequently examined the
sensitivity of sentence CNNs to hyperparameter choices and
stochastic training, highlighting the importance of reporting
variation across repeated runs.
The published TREC results provide context for model selection;
they are not directly comparable with results on the restricted
binary subset used in this study.

\subsection{Quantum and Hybrid Methods for NLP Classification}

Grammar-based quantum natural language processing builds on
compositional models that combine word meanings according to
grammatical structure \cite{coecke2010}.
Meichanetzidis et al. \cite{meichanetzidis2023} mapped sentences
to parameterised quantum circuits and implemented a supervised
binary sentence classification task on quantum hardware.
This demonstrated the feasibility of representing and training
a grammar-aware NLP model on a small quantum processor.
Lorenz et al. \cite{lorenz2023} extended experimental QNLP to
datasets containing more than one hundred sentences and compared
syntax-sensitive models with alternatives using less or no syntax.
They reported convergence in both classical simulations and
quantum hardware experiments, with the usefulness of syntactic
structure depending on the classification task.

Metawei et al. \cite{metawei2023} investigated whether pairs of
sentences belonged to the same topic using a hybrid QNLP model.
Their experiments reported accuracy improvements over a classical
tensor-network baseline and examined how convergence changed with
problem size, circuit ansatz and simulated noise.
These studies demonstrate quantum NLP classification under
specific, small-scale experimental conditions, using tasks and
datasets distinct from TREC question classification.

A complementary approach combines classical feature extraction
with quantum processing.
Mari et al. \cite{mari2020} introduced dressed quantum circuits,
in which a pretrained classical network supplies features to
a variational quantum circuit followed by a classical output layer.
Their transfer-learning experiments demonstrated that compact
quantum layers could be integrated into classical learning pipelines.
For text classification, Yang et al. \cite{yang2022} combined a
pretrained BERT encoder with quantum temporal convolution.
They reported competitive intent classification results on the
Snips and ATIS datasets.
Together, these studies illustrate two approaches to quantum NLP
classification: constructing circuits from linguistic structure,
and processing classical text representations with quantum layers.

\subsection{Computational Efficiency of Quantum Simulation}

Gradient computation is an important part of the computational
cost of variational circuit training.
Mitarai et al. \cite{mitarai2018} developed quantum circuit learning
with gradients obtained from circuit evaluations.
Schuld et al. \cite{schuld2019} established analytic gradient
evaluation methods, including parameter-shift rules that recover
derivatives from circuits evaluated at shifted parameter values.
Wierichs et al. \cite{wierichs2022} extended this framework through
generalised parameter-shift rules.
These methods enable gradient-based optimisation, but introduce
additional circuit evaluations into the training workflow.

Simulation frameworks also address the cost of differentiation
and circuit execution.
Luo et al. \cite{luo2020} introduced Yao.jl, a differentiable
quantum simulation framework featuring an automatic differentiation
engine optimised for reversible computing and GPU-supported
batched quantum registers.
Their work demonstrates that differentiation and batching can
be incorporated into the design of an efficient simulator.
Doi and Horii \cite{doi2020} addressed data movement in
large-scale distributed simulation by implementing cache blocking
in Qiskit Aer.
They evaluated the implementation on GPU clusters and demonstrated
scalability for large quantum circuit simulations.

These studies address complementary aspects of computational
efficiency: gradient evaluation, simulator differentiation,
batching and distributed state-vector execution.
Their findings motivate assessing acceleration within the
complete training workflow.
In particular, the performance of large-circuit simulation
does not by itself determine the runtime of a hybrid classifier
that repeatedly evaluates a small circuit during training.
This study examines that setting through a four-qubit CNN--VQC
pipeline.

\section{Methodology}
\label{sec:method}
{\small
\begin{center}
Question $\rightarrow$ word embeddings $\rightarrow$ CNN features $\rightarrow$ four rotation angles $\rightarrow$ four-qubit circuit $\rightarrow$ two-class prediction
\end{center}
}
\begin{figure}[h!]
\centering
\includegraphics[width=\linewidth]{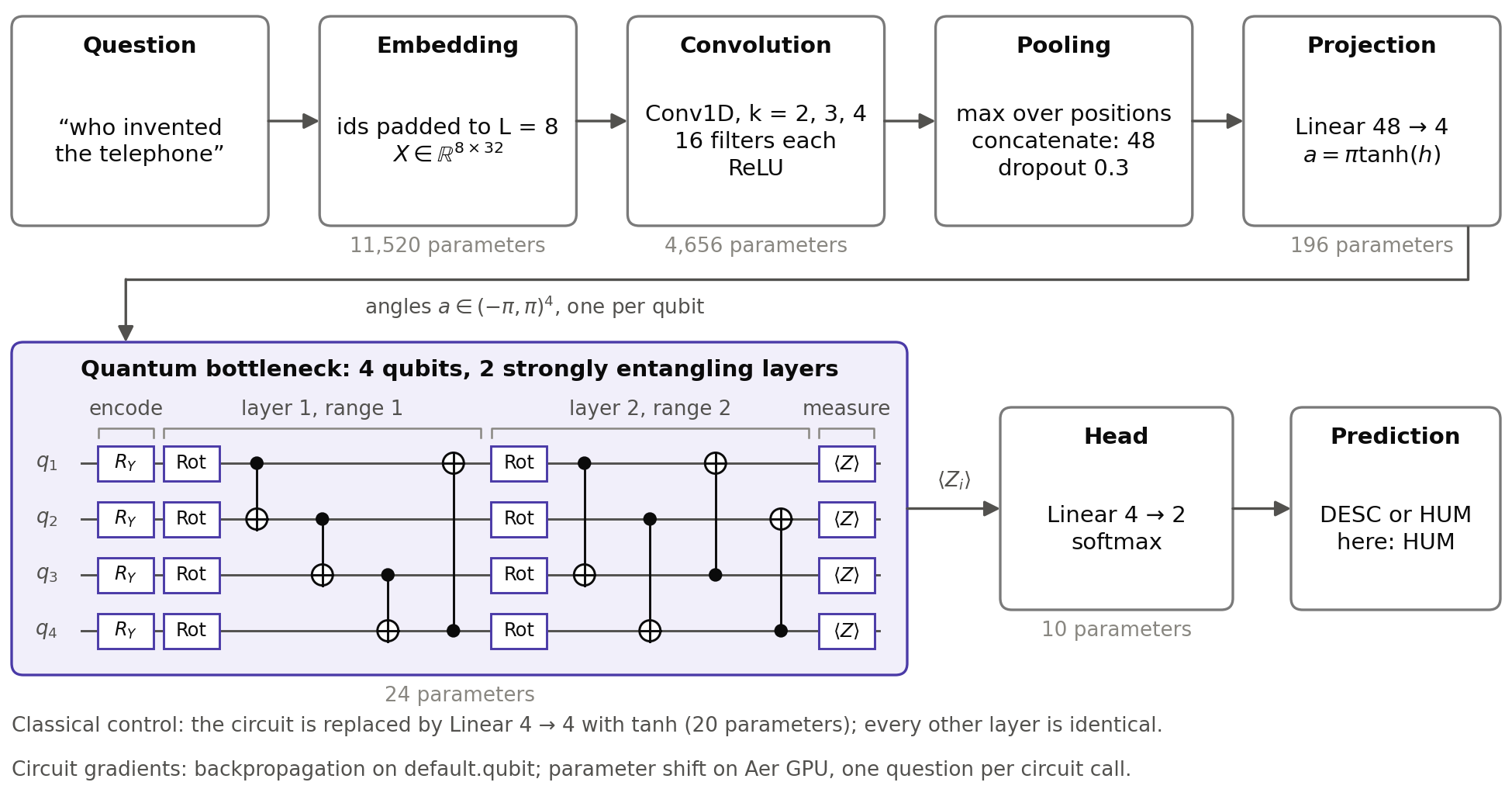}
\caption{The CNN and VQC classifier, with the parameter count of each trainable stage. Row one is the classical front end and row two the quantum bottleneck and classification head. In the circuit every qubit is first rotated by its angle $a_i$. Each strongly entangling layer then applies a general rotation to every qubit, followed by a CNOT from qubit $i$ to qubit $i+r$ (modulo 4), with range $r=1$ in layer one and $r=2$ in layer two. The example question is taken from the training split.}
\label{fig:architecture}
\end{figure}
The proposed pipeline combines classical text feature extraction with a variational quantum circuit for binary question classification. Questions are first tokenised and mapped to trainable word embeddings. A TextCNN extracts a fixed-dimensional feature representation, which is projected to four rotation angles for encoding into a four-qubit circuit. The circuit produces four expectation values, and a classical linear layer maps these outputs to the two class logits. The classical and quantum components are trained jointly. Figure 1 illustrates the model architecture.

The classical control keeps this network unchanged except for the bottleneck: it replaces the four-qubit circuit (24 parameters) with a similarly sized classical layer, a linear map from the four angles to four outputs followed by a hyperbolic tangent (20 parameters). The comparison is therefore a focused test of the four-dimensional bottleneck, not a comparison with a strong general-purpose text classifier.

\subsection{Dataset}
\label{sec:dataset}
We used the TREC question classification dataset
\cite{li2002, trec_huggingface}. It contains 5,452 training and 500 test questions labelled with six coarse answer types: abbreviation (ABBR), description (DESC), entity (ENTY), human (HUM), location (LOC) and numeric value (NUM). The coarse classes are unevenly represented in the training file, which holds 1,250 ENTY, 1,223 HUM, 1,162 DESC, 896 NUM, 835 LOC and 86 ABBR questions. Training questions average about ten whitespace-delimited tokens, with a maximum of 37, over a vocabulary of about 8,700 words. 

\subsection{Sampling Criteria}
\label{sec:sampling}
The binary subset was first built for a grammar-based quantum model whose circuit width grows with sentence length and whose parameter count grows with vocabulary size. We reused it unchanged, so its length and vocabulary limits are stricter than the CNN and VQC model requires. The procedure, implemented in a single script with a fixed seed of 0, applied eight steps. Table~\ref{tab:sampling} gives the number of questions that survived each one.

\begin{algorithm}[t]
\caption{Q\&A Dataset Preprocessing and Vocabulary-Aware Splitting}
\label{alg:dataset_preprocessing}
\begin{algorithmic}[1]
\Require Official training and test datasets
\Ensure Balanced training and test sets with an 80:20 ratio

\State Merge the official training and test datasets into a single pool of 5,952 questions.
\ForAll{questions $q$ in the pooled dataset}
    \State Convert $q$ to lowercase.
    \State Remove punctuation-only tokens from $q$.
\EndFor

\State Retain questions containing at most eight words.
\State Select only DESC and HUM questions and remap their labels to 0 and 1, respectively.
\State Remove repeated question texts, keeping the first occurrence of each.

\State Count the occurrences of every word across the remaining questions.
\State Remove every question containing a word that occurs fewer than two times, using these counts without recounting.

\State Randomly downsample the majority class to the size of the minority class, and call the result $D$.

\State Let $V(X)$ be the set of words that occur in the questions $X$.
\State Initialize training set $D_{\mathrm{train}} \gets \emptyset$ and test set $D_{\mathrm{test}} \gets \emptyset$.
\ForAll{classes $c$ in $\{\mathrm{DESC}, \mathrm{HUM}\}$}
    \State Shuffle the questions of class $c$ in $D$, giving $D_c$.
    \State Set the class quota $k_c$ to the nearest integer to $0.2|D_c|$.
    \ForAll{questions $q$ in $D_c$}
        \If{$|D_{\mathrm{test}} \cap D_c| < k_c$ \textbf{and}
            $\forall w \in q,\; w \in V(D \setminus (D_{\mathrm{test}} \cup \{q\}))$}
            \State Add $q$ to $D_{\mathrm{test}}$.
        \Else
            \State Add $q$ to $D_{\mathrm{train}}$.
        \EndIf
    \EndFor
\EndFor

\State \Return $D_{\mathrm{train}}, D_{\mathrm{test}}$
\end{algorithmic}
\end{algorithm}

\begin{table}[htbp]
\centering
\caption{Questions remaining after each sampling step. The first row counts all six classes. The DESC and HUM columns count those two classes at every step.}
\label{tab:sampling}
\begin{tabular}{lrrr}
\toprule
\textbf{Step} & \textbf{Questions} & \textbf{DESC} & \textbf{HUM} \\
\midrule
Official training and test files pooled & 5,952 & 1,300 & 1,288 \\
Cleaned; at most 8 words; DESC or HUM & 1,468 & 921 & 547 \\
Duplicate texts removed & 1,448 & 906 & 542 \\
Every word occurs at least twice & 224 & 115 & 109 \\
Classes balanced & 218 & 109 & 109 \\
\midrule
Training split & 174 & 87 & 87 \\
Test split & 44 & 22 & 22 \\
\bottomrule
\end{tabular}
\end{table}

We re-split the pooled TREC files to retain the vocabulary constraint of the subset originally constructed for the grammar-based model. After the length and frequency filters, almost no official test question consisted only of words that remained in the filtered training questions. This constraint is not required by the CNN and VQC architecture, which can map unseen words to an unknown-word token, at the cost of losing their lexical identity. Under the pooled and vocabulary-aware split, no test question contains a word absent from the training split, and no question text appears in both splits. As a consequence, 21 of the 174 training questions and 2 of the 44 test questions come from the official test file, so accuracies on this subset are not comparable with published results on the official TREC test set. The evaluation is restricted to short questions whose words occur in training; excluding duplicate texts does not remove this selection bias. Training questions contain 5.8 words on average (range 2 to 8) and test questions 5.3 (range 3 to 8).

\subsection{Tokenisation and Embedding}
\label{sec:embedding}
Each question was split into lowercase tokens matching the regular expression \texttt{[a-z0-9']+}. The vocabulary was built from the training split alone and holds its 358 distinct tokens together with a padding token and an unknown-word token, 360 entries in total. Because of the vocabulary-aware split, the unknown-word token never occurs in the test set. Each question became a sequence of token indices, padded with the padding index to the length $L=8$ of the longest question. An embedding layer with $d=32$ dimensions maps the sequence to a matrix
\begin{equation}
    X \in \mathbb{R}^{L \times d}, \qquad L = 8,\; d = 32,
\end{equation}
whose rows are the word vectors. The word vectors were initialized randomly and learned during training, and the padding row was fixed at zero.

\subsection{CNN Feature Extractor}
\label{sec:cnn}
Our research draws inspiration from the framework proposed by kim et al.\cite{kim2014}, three one-dimensional convolutions slide over the sequence with kernel widths $k \in \{2, 3, 4\}$ words and 16 filters each. Each convolution pads the sequence by $k-1$ positions on both sides, so questions shorter than a kernel still produce outputs, and is followed by a rectified linear unit. Taking the maximum of every filter over all positions and concatenating the three results gives
\begin{equation}
    c_k = \mathrm{maxpool}\big(\mathrm{ReLU}(\mathrm{Conv}_k X)\big) \in \mathbb{R}^{16}, \qquad c = [\, c_2,\; c_3,\; c_4 \,] \in \mathbb{R}^{48}.
\end{equation}
Dropout with rate 0.3 \cite{srivastava2014} is applied to $c$ during training. A linear projection then reduces the 48 features to one value per qubit, and a scaled hyperbolic tangent turns each value into a rotation angle:
\begin{equation}
    h = W c + b \in \mathbb{R}^{4}, \qquad a = \pi \tanh(h) \in (-\pi, \pi)^{4}.
\end{equation}
The scaling keeps every angle within one period of the rotation. Without it, two very different projections could wrap around to the same rotation and become indistinguishable to the circuit.

\subsection{Quantum Classifier}
\label{sec:vqc}
The circuit acts on four qubits initialized in $|0\rangle$. Angle encoding rotates the qubit $i$ about the $y$ axis by $a_i$,
\begin{equation}
    |\psi(a)\rangle = \bigotimes_{i=1}^{4} R_Y(a_i)\,|0\rangle .
\end{equation}
Two strongly entangling layers follow \cite{schuld2020}. Layer $\ell$ applies to every qubit a general single-qubit rotation
\begin{equation}
    \mathrm{Rot}(\phi,\vartheta,\omega) = R_Z(\omega)\,R_Y(\vartheta)\,R_Z(\phi)
\end{equation}
and then, for $i = 1, \dots, 4$ in turn, a CNOT with control qubit $i$ and target qubit $i + r_\ell$ modulo 4, where the range is $r_1 = 1$ in the first layer and $r_2 = 2$ in the second. Writing the two layers as $U(\theta)$, with $\theta \in \mathbb{R}^{2 \times 4 \times 3}$ holding 24 trainable angles, the circuit returns the four Pauli-$Z$ expectation values
\begin{equation}
    z_i = \langle \psi(a) |\, U(\theta)^{\dagger} Z_i\, U(\theta) \,| \psi(a) \rangle \in [-1, 1], \qquad i = 1, \dots, 4 .
\end{equation}
The circuit weights were initialized uniformly in $[0, 2\pi)$. Both simulators computed $z$ exactly from the state vector, without sampling measurement shots. A linear head $V z + v$ with $V \in \mathbb{R}^{2 \times 4}$ maps the expectation values to two logits, and a softmax gives the class probabilities. The circuit was implemented as a PennyLane layer \cite{pennylane} inside a PyTorch model \cite{paszke2019}. With 11,520 embedding, 4,656 convolution, 196 projection, 24 circuit and 10 head parameters, the hybrid holds 16,406 trainable parameters, of which the circuit holds 0.15\%.

\subsection{Classical Control}
\label{sec:control}
The classical control is the same network with the circuit replaced by a linear map from the four angles to four outputs followed by a hyperbolic tangent, $z = \tanh(A a + e)$ with $A \in \mathbb{R}^{4 \times 4}$ and $e \in \mathbb{R}^{4}$, which has 20 parameters. Its input $a = \pi \tanh(h)$ is computed exactly as the circuit's input angles (Section~\ref{sec:cnn} and Figure~\ref{fig:architecture}), so the control applies a second hyperbolic tangent after the angle scaling rather than replacing it. This second hyperbolic tangent keeps the control's bottleneck in $[-1, 1]$, the range of the expectation values it replaces. Everything else, including the embedding, the convolutions, the projection, the angle scaling and the head, is identical, and the control holds 16,402 parameters. The comparison therefore asks whether the four-dimensional bottleneck is better computed by a small circuit or by a small classical layer.

\subsection{End-to-End Training}
\label{sec:training}
All models minimised the cross-entropy loss with the Adam optimiser \cite{kingma2015}, a learning rate of $5 \times 10^{-3}$ and an $L_2$ weight decay of $10^{-5}$, for 20 epochs with the training set reshuffled every epoch. The PyTorch and NumPy random seeds were set to 0 before each model was built, so all models started from the same convolutional weights. Gradients with respect to the classical parameters were computed by backpropagation. The circuit contributes gradients both for its own 24 weights and for the four input angles, through which the loss reaches the convolutional front end.

How these circuit gradients were obtained depended on the simulator. The \texttt{default.qubit} simulator in PennyLane is written in the same automatic-differentiation framework as the model, so it backpropagated through the simulated state vector and processed a whole batch of questions in one call. Qiskit Aer, accessed through the PennyLane--Qiskit plugin \cite{qiskitaer,pennylaneqiskit}, offers no such path, and PennyLane therefore used the parameter-shift rule
\begin{equation}
    \frac{\partial z_i}{\partial \theta_j} = \frac{1}{2}\left[ z_i\!\left(\theta + \tfrac{\pi}{2} e_j\right) - z_i\!\left(\theta - \tfrac{\pi}{2} e_j\right) \right],
    \label{eq:shift}
\end{equation}
where $e_j$ is the $j$-th unit vector \cite{mitarai2018,schuld2019}. Every angle in the circuit enters a Pauli rotation, so the rule is exact and needs two evaluations per angle \cite{wierichs2022}. PennyLane's parameter-shift implementation cannot differentiate a batch of inputs processed in one call, so each question in a batch was executed and differentiated separately. For each question, the rule requires $2 \times (24 + 4) = 56$ shifted circuit executions in addition to the forward execution. In exact arithmetic, state-vector backpropagation and the parameter-shift rule yield equivalent gradients for the same circuit and parameter values. This mathematical equivalence alone does not establish identical numerical training trajectories across implementations, and the training schedules differed as described in Section~\ref{sec:setup}. The observed differences between the training runs therefore cannot be attributed solely to the training schedule.

\section{Experimental Setup}
\label{sec:setup}
This section presents the hardware and software environments used to conduct the experiments, details the three full training configurations and the CPU-based projection approach, and describes the evaluation protocol and runtime measurement methodology.
\subsection{Hardware and Software}
The three full-training configurations were executed within a
single job on a commercial HPC node equipped within the following test environment:

\begin{table}[htbp]
\centering
\small
\caption{Experimental Environment (Hardware and Software) }
\label{tab:configs}
\begin{tabular}{llllr}
\toprule
\textbf{Component} & \textbf{Configuration}  \\
\midrule
GPU Core & NVIDIA Tesla V100S-PCIE-32GB \\
CPU Core & 32 virtual CPUs - system Memory Size 128\,GB \\
Operating System & Linux, kernel 7.2.2 \\
PennyLane & 0.40 \\
Qiskit Aer CPU-based Simulator & 0.17.2 \\
Qiskit Aer GPU-based Simulator & 0.15.1 (CUDA 12) \\
Other packages & Python 3.12, CUDA 12.6 and NVIDIA
driver 560.x.\\

\bottomrule
\end{tabular}

\end{table}

Exact package versions were not recorded with the runs.
The training code states that the GPU server used PennyLane 0.40;
the versions of PyTorch, the PennyLane--Qiskit plugin and
Qiskit Aer are not known.
The classical components of every model, including the embedding,
convolutional feature extractor, projection and classification head,
were executed on the CPU.
Only the quantum state-vector simulation in the Aer GPU
configuration used the GPU.
Aer was configured to use double-precision state-vector simulation.
Both hybrid configurations computed expectation values directly
from the state vector without measurement-shot sampling.

Additional Aer CPU timing measurements were collected on a
separate CPU-only machine.
These measurements were used to estimate training cost;
they did not constitute a complete training run.

\subsection{Configurations}
Table~\ref{tab:configs} lists the configurations. The three full runs differ in more than the device. First, batch size: the classical CNN and the hybrid on \texttt{default.qubit} were trained with batches of 16 questions, and the hybrid on Aer with batches of 4, so it took 44 optimisation steps per epoch against 11 for the other two. The learning rate was the same in all runs, so the Aer run also made four times as many parameter updates. Second, gradient method: the \texttt{default.qubit} hybrid backpropagated through the simulated state for a whole batch at once, whereas the Aer hybrid used the parameter-shift rule one question at a time (Section~\ref{sec:training}). Third, device: only Aer's state-vector simulation ran on the GPU, and the embedding, convolutions, projection and head of every model ran on the CPU. The learning rate will be further discussed in Section~\ref{sec:limitations}.

\begin{table}[htbp]
\centering
\small
\caption{Training configurations. All three full runs used 20 epochs, seed 0, Adam with learning rate $5 \times 10^{-3}$ and exact (shot-free) simulation. }
\label{tab:configs}
\begin{tabular}{llllr}
\toprule
\textbf{Configuration} & \textbf{Bottleneck} & \textbf{Circuit gradient} & \textbf{Device} & \textbf{Batch} \\
\midrule
Classical CNN & Linear $4 \to 4$, $\tanh$ & n.a. & CPU & 16 \\
Hybrid, \texttt{default.qubit} & 4-qubit VQC & backpropagation, batched & CPU & 16 \\
Hybrid, Aer GPU & 4-qubit VQC & parameter shift, per question & GPU & 4 \\
Hybrid, Aer CPU (projection) & 4-qubit VQC & parameter shift, per question & CPU$^\dagger$ & 2 \\
\bottomrule
\end{tabular}
\begin{tablenotes}
\small
\item $^\dagger$The CPU projection used the same model and gradient method on a different, CPU-only machine; it timed three optimisation steps and did not train the model.
\end{tablenotes}
\end{table}

\subsection{Protocol and Metrics}
Every configuration was trained once, with seed 0. No hyperparameter search was performed: every run used the training script's default settings, which were not changed before the runs, except that the number of epochs was lowered from the default of 30 to 20. Before that change, the classical control and the \texttt{default.qubit} hybrid had been trained for 30 epochs on the development workstation and scored on the same test set; no reason for the change was recorded. After these runs, a smaller configuration (16-dimensional embeddings, 8 filters per kernel, a one-layer circuit, 100 epochs) was started on the GPU node; it did not finish and is not reported. There was no validation split, no early stopping and no checkpoint selection. After each epoch the model was evaluated, with dropout disabled, on the full training and test sets. The per-epoch test accuracy was recorded for monitoring only and did not influence training, and the reported accuracy is that after the twentieth epoch. We report accuracy only. The test set holds 22 questions of each class, so accuracy equals the mean recall over the two classes. Per-question predictions were not stored, so neither a confusion matrix nor a per-class F1-score is available.

Training time was measured as wall-clock time around each configuration's training loop. It includes the per-epoch evaluation passes and excludes data loading and model construction. The whole job, all three configurations included, was also timed by the operating system.

The Aer CPU cost was not measured on the HPC node. It comes from a separate run on a CPU-only development workstation, in which three optimization steps on batches of two questions took 15.72\,s per step on average, or 7.86\,s per question. Multiplying by the 174 training questions gives a projected 1,368\,s per epoch. The projection excludes evaluation passes and assumes a constant cost per question.

\section{Results}
\label{sec:results}
This section presents the test accuracy and training convergence of the three full experimental configurations, followed by a comparative analysis of their training times and an estimate of the computational cost of executing Aer-based training on a CPU.
\subsection{Accuracy}
Table~\ref{tab:results} reports test accuracy and training time. The hybrid trained on Aer classified 43 of the 44 test questions correctly, the classical CNN 42 and the hybrid trained on \texttt{default.qubit} 40. The Aer hybrid therefore exceeded the classical CNN by one question, or 2.27 percentage points; the difference of the rounded percentages in Table~\ref{tab:results} is 2.28 points. The \texttt{default.qubit} hybrid trailed the classical CNN by two questions. Both hybrid runs started from the same initial weights and optimised a model with the same outputs and gradients (Section~\ref{sec:training}). They differed in batch size and therefore in the order and number of updates, and they ended three test questions apart. In all three runs the final test accuracy was also the highest reached at any epoch.

\begin{table}[htbp]
\centering
\small
\caption{Accuracy after 20 epochs and training time. Each test question is worth 2.27 percentage points. Times include per-epoch evaluation, except in the projected row, which is an extrapolation from three optimisation steps on a different machine, not a training run; its ratio to the Aer GPU row is therefore indicative only.}
\label{tab:results}
\begin{tabular}{lrrrrr}
\toprule
\textbf{Configuration} & \textbf{Correct} & \textbf{Test acc.} & \textbf{Train acc.} & \textbf{Total time} & \textbf{Per epoch} \\
\midrule
Classical CNN & 42 / 44 & 95.45\% & 100.00\% & 1.3\,s & 0.065\,s \\
Hybrid, \texttt{default.qubit} & 40 / 44 & 90.91\% & 99.43\% & 4.8\,s & 0.242\,s \\
Hybrid, Aer GPU & 43 / 44 & 97.73\% & 99.43\% & 10,127\,s & 506.4\,s \\
Hybrid, Aer CPU (projected) & n.a. & n.a. & n.a. & n.a. & 1,368\,s \\
\bottomrule
\end{tabular}
\end{table}

\begin{figure}[htbp]
\centering
\includegraphics[width=\linewidth]{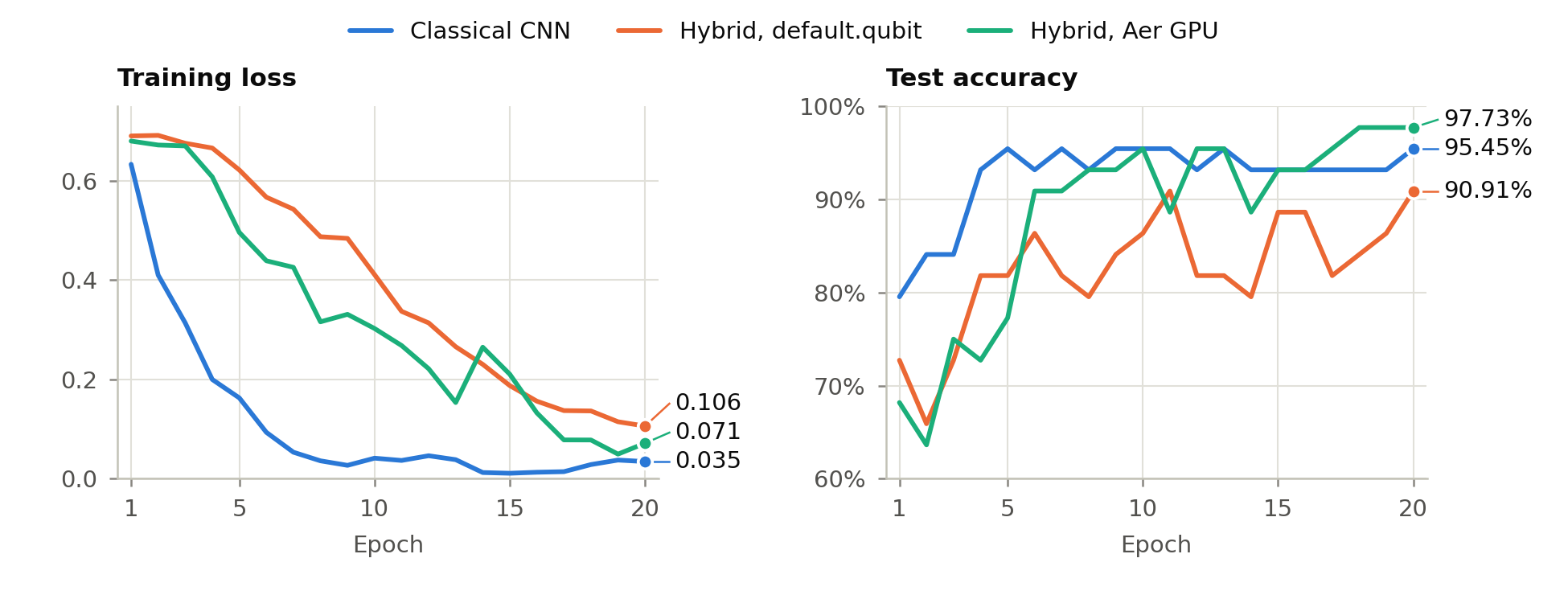}
\caption{Training loss and test accuracy after every epoch, for one seed. Labels give the values after epoch 20. One test question corresponds to 2.27 percentage points, which is the smallest step the accuracy curves can take.}
\label{fig:curves}
\end{figure}

Figure~\ref{fig:curves} shows how training progressed. The classical CNN fitted the training set fastest: its training accuracy reached 100\% at epoch 5, its training loss stayed below 0.05 from epoch 8 onward, and its test accuracy stayed between 93.18\% and 95.45\% from epoch 4 onward. Both hybrids reduced the loss more slowly, from about 0.69 to 0.106 on \texttt{default.qubit} and from 0.680 to 0.071 on Aer. The Aer hybrid first exceeded 90\% test accuracy at epoch 6, varied between 88.64\% and 95.45\% until epoch 17, and held 97.73\% over the last three epochs. The \texttt{default.qubit} hybrid varied between 79.55\% and 90.91\% after epoch 4. Gradients reached the circuit in both hybrid runs: at the final batch the Euclidean norm of the circuit-weight gradient was 0.175 on \texttt{default.qubit} and 0.013 on Aer.

\subsection{Training Cost}
\begin{figure}[htbp]
\centering
\includegraphics[width=\linewidth]{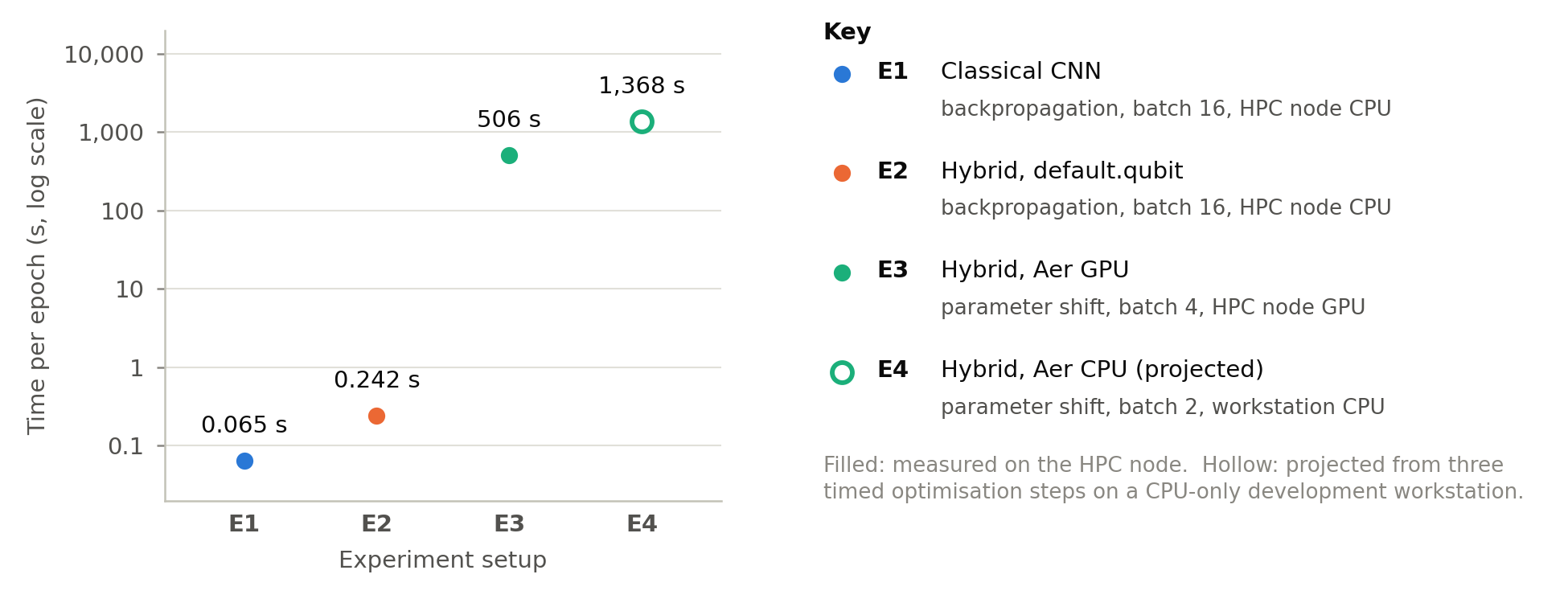}
\caption{Training time per epoch on a logarithmic scale. Filled markers are measured on the HPC node; the hollow marker is the projection from three timed optimisation steps on a CPU-only development workstation. Setups E1--E4 are the rows of Table~\ref{tab:configs}, in order.}
\label{fig:cost}
\end{figure}

Runtime-based training cost revealed substantially greater differences among configurations than accuracy, as illustrated in Figure \ref{fig:cost}. The classical CNN trained in 1.3\,s. The hybrid on \texttt{default.qubit} took 4.8\,s, about four times as long. The hybrid on Aer GPU took 10,127\,s, or 2 hours 49 minutes: about 2,100 times the cost of the same model on \texttt{default.qubit} and about 7,800 times that of the classical CNN. 
Its cost per epoch was nearly constant, between 501.5\,s and 509.3\,s over the 20 epochs, or about 2.9\,s per training question including evaluation. Against the CPU projection of 1,368\,s per epoch, the GPU run took 2.70 times less time per epoch. This is an indicative comparison, not a measured GPU speedup: the CPU value was projected from three optimisation steps on a different machine and excludes the evaluation passes included in the GPU timing.

The operating-system timing of the whole job recorded 10,140\,s of wall-clock time and 10,093\,s of user CPU time, so the host process kept one CPU core busy for practically the entire run. Its peak resident memory was 1.2\,GB of the 128\,GB available.

\section{Discussion}
\label{sec:discussion}
Referring to the research questions introduced in Section~\ref{sec:intro}, we discuss the classification performance of the hybrid model and the training cost of the different execution configurations.

\subsection{Training Cost and Gradient Computation}

The large runtime difference between the two hybrid configurations highlights the importance of gradient computation and simulator integration. A four-qubit state vector holds 16 amplitudes, and applying the circuit's gates to it is a negligible computation on either a CPU or a GPU. Yet the same model trained about three orders of magnitude more slowly on Aer GPU than on \texttt{default.qubit}. The two runs differed in how circuit gradients were obtained. On \texttt{default.qubit}, backpropagation through the simulated state handled a whole batch of questions in one differentiable call. On Aer, the parameter-shift rule of Eq.~\eqref{eq:shift} required 57 circuit executions for every training question at every step, and the questions could not be batched. The number of executions per epoch, not the work inside each execution, therefore likely set the training time. The timing pattern agrees with this reading: the cost per epoch barely changed across epochs, memory use stayed small, and the host process kept one CPU core busy throughout. We did not profile individual executions, and a busy host core is also consistent with a process waiting on the GPU, so this pattern supports the reading without proving it.

\subsection{GPU Comparison and Its Limitations}
The GPU run was faster than the CPU projection, which indicates, but does not measure, that Aer's GPU device reduced cost even for a circuit this small. The GPU runtime was measured on the HPC node, whereas the CPU estimate was extrapolated from short timing measurements on a separate CPU-only machine. GPU state-vector simulation is built to accelerate operations on large states \cite{doi2020}. With four qubits there is little work inside a single execution to parallelise, while every execution still pays the fixed cost of a call. On this workload the benefit of HPC hardware therefore did not follow automatically from the hardware. It likely depended on three factors: batching, meaning whether questions and shifted circuits can be dispatched together; the gradient method, meaning whether the number of executions grows with the number of parameters; and simulator integration, meaning whether the simulator takes part in automatic differentiation. Together, these factors mattered far more than the device. Backpropagation on a CPU was faster than parameter shift on a GPU by a much larger factor than the GPU gained over the CPU under parameter shift.

\subsection{Representation and resource control} 
The fixed-width design controlled quantum resources as intended: every question used four qubits and 24 circuit parameters, whatever its length. The cost of that control is that the circuit sees only four numbers per question, chosen by the learned projection. Almost all of the model's parameters sit in the embedding and the convolutions, and the circuit holds a small fraction of one percent. A model that encodes each input once as a rotation angle can represent only low-degree Fourier series in that input \cite{schuld2021}, so the circuit's function class on its four inputs is narrow. It is therefore unsurprising that a 20-parameter classical layer could take its place. On this task the front end CNN, not the bottleneck, appears to determine most of what the model learns.

\subsection{Trainability} 
Both hybrid configurations, the one trained on CPU-based default.qubit simulation and Aer GPU simulation, supported end-to-end optimisation, as indicated by decreasing training loss, high final training accuracy and non-zero circuit-weight gradients. End-to-end optimisation through a four-qubit bottleneck was therefore feasible. Both hybrids reduced the training loss more slowly than the classical control. At four qubits and two layers, the vanishing gradients that affect deep random circuits \cite{mcclean2018} are an unlikely cause. The current experiments do not isolate the causes of the slower loss reduction in the hybrid models. Further experiments are needed to examine the effects of angle encoding, circuit architecture and optimisation settings.

\subsection{Accuracy} 
Both hybrid configurations, the one trained on CPU-based default.qubit simulation and Aer GPU simulation, used the same architecture and initial weights, but differed in batch size, backend and gradient implementation. This gap was larger than the one-question difference between the Aer hybrid and the classical control. The present comparison cannot disentangle the effect of the quantum bottleneck from that of the training configuration. What the results do establish is feasibility: a CNN and VQC hybrid trained end to end on a GPU simulator reached test accuracy in the same range as its classical control on this subset.

\section{Limitations}
\label{sec:limitations}
This section discusses four key limitations of the study: the use of a single random seed, the restricted dataset scope and test-set size, variations in training and timing conditions across experimental runs, and the absence of detailed profiling to identify the primary sources of computational overhead.

\subsection{Single-Seed Evaluation}

Each of the three full-training configurations was evaluated
using a single random seed.
Previous studies have documented variation in neural model
performance due to stochastic training
\cite{zhang2017,reimers2017}.
The reported accuracies are therefore outcomes of individual
runs and do not characterise variability across random seeds.
The dataset sampling procedure also used a fixed seed,
so sensitivity to the selection of training and test questions
was not evaluated.

\subsection{Dataset Scope and Test-Set Size}

The evaluation used a balanced binary subset of TREC containing
short questions whose words occurred in the training split.
The subset was originally prepared for a grammar-based model
and was obtained by pooling and re-splitting the official files.
The reported accuracies are therefore specific to this subset
and cannot be directly compared with published results on the
official TREC test set.
Performance on the full six-class task, longer questions,
unseen vocabulary and other datasets remains untested.

The test set contained 44 questions, so one correct classification
corresponded to approximately 2.27 percentage points.
The accuracy difference between the Aer GPU hybrid and the
classical CNN amounted to one question.
This small test set limits the resolution of the comparison.
Per-question predictions were not stored, preventing paired
statistical analysis such as McNemar's test
\cite{dietterich1998}.
The observed accuracy difference does not establish a consistent
performance improvement.

No validation split was used.
Test accuracy was monitored after each epoch, but the reported
result was taken from the final epoch without checkpoint selection.
No hyperparameter search was performed, and only the number of
epochs differed from the script defaults (Section~\ref{sec:setup}).
Because 30-epoch runs had already been scored on the same test set
before that change, the test set was not untouched when the epoch
count was chosen.

\subsection{Differences in Training and Timing Conditions}

The Aer GPU hybrid used a batch size of 4, whereas the classical
CNN and the \texttt{default.qubit} hybrid used a batch size of 16.
With the same number of epochs and learning rate, the Aer GPU
configuration performed four times as many parameter updates.
The two hybrid configurations also differed in simulation backend
and gradient computation method.
Consequently, the accuracy differences cannot be attributed
solely to the quantum bottleneck or the execution device.

The Aer CPU runtime was projected from three optimisation steps
with batches of two questions on a separate CPU-only machine.
The projection assumed a constant cost per question and excluded
evaluation passes, whereas the measured GPU runtime included them.
The reported ratio of approximately 2.70 is therefore an indicative
comparison between these configurations, rather than a matched
measurement of GPU acceleration.

\subsection{Training Cost and Bottleneck Identification}

The Aer GPU configuration required substantially more training
time than both the classical CNN and the
\texttt{default.qubit} hybrid.
This cost constrains repeated evaluation and experiments on
larger datasets under the implemented workflow.
It should not be interpreted as an intrinsic cost of the
hybrid architecture, since the state-vector backpropagation
configuration trained much faster.

Individual circuit executions were not profiled, and GPU
utilisation was not recorded.
The measurements therefore do not separate gradient evaluation,
simulator interface overhead, data transfer and device computation.
Further profiling is needed to identify their respective
contributions to the observed runtime and guide optimisation.

\section{Conclusion and Future Work}
\label{sec:conclusion}

This study presented and evaluated a hybrid CNN--VQC pipeline
for binary question classification on a balanced TREC subset.
The model combined trainable classical text features with a
fixed-width four-qubit circuit and was trained end to end
using two simulation configurations.
The Aer GPU hybrid achieved 97.73\% test accuracy, compared
with 95.45\% for the classical CNN baseline and 90.91\% for
the hybrid using CPU-based \texttt{default.qubit} simulation.
The Aer GPU result exceeded the classical baseline by one
correctly classified question, demonstrating the feasibility
of the hybrid pipeline on the selected task without establishing
a consistent accuracy benefit from the quantum layer.

Training runtime differed substantially across configurations.
The classical CNN required 1.3\,s, the
\texttt{default.qubit} hybrid required 4.8\,s, and the
Aer GPU hybrid required 10,127\,s.
The measured Aer GPU runtime was lower than the projected
Aer CPU training time, but remained substantially higher
than that of the CPU hybrid using state-vector backpropagation.
These results highlight the importance of evaluating GPU
acceleration within the complete hybrid training workflow,
including the simulator, gradient computation method and
batching strategy.
Repeated parameter-shift evaluations likely contributed to the
Aer training overhead, but profiling is needed to identify the
dominant source.

Future work will extend the evaluation through repeated runs,
a validation split, stored per-question predictions and
uncertainty estimates.
Comparisons will use consistent batch sizes and training budgets,
with CPU/GPU timing conducted under matched simulator,
gradient computation and measurement settings.
Training efficiency will be investigated through GPU-compatible
implementations of simulator-based differentiation
\cite{luo2020} and stochastic gradient estimation methods
such as simultaneous perturbation, which uses two loss
evaluations per perturbation \cite{spall1992}.
Profiling circuit execution and recording GPU utilisation
will help distinguish device computation from interface,
scheduling and data-transfer overhead.

Further experiments will evaluate the full six-class TREC task
and extend preprocessing to accommodate longer questions
while retaining a fixed-width quantum layer.
Ablation studies using different classical feature extractors
and quantum or classical bottlenecks will assess the contribution
of each component.
Finite-shot and noisy simulations will then examine the
robustness and computational requirements of the pipeline
before evaluation on quantum hardware.

\section*{CRediT authorship contribution statement}

\textbf{Sk Saif Ibna Ezhar Arko:} Conceptualization, Methodology, Formal analysis, Data curation, Investigation, Writing -- original draft, Visualization. 
\textbf{Zijia Zhao:} Investigation, Resources, Validation, Visualization, Writing -- original draft, Writing -- review \& editing. 
\textbf{Maha A. Metawei:} Writing -- review \& editing, Validation, Supervision, Project administration.

\section*{Ethics statement}

This study used the publicly available TREC question classification dataset and did not involve direct experimentation, interaction, or intervention with human participants or animals. No personally identifiable information was collected or processed. Therefore, ethical approval and informed consent were not required. The study was conducted using publicly available data for research purposes.

\section*{AI Disclosure}
The authors utilized AI-assisted tools to enhance language polishing and readability improvements. All content was reviewed and verified by the authors to ensure accuracy and integrity.

\section*{Funding statement}
This study received no particular grants from public, commercial, or non-profit funding entities.

\section*{Declaration of competing interest}
The authors declare that they have no known competing financial interests or personal relationships that could have appeared to influence the work reported in this paper.

\section*{Acknowledgments}
This work was carried out within the QIntern 2026 project ``Accelerating hybrid quantum-classical machine learning tasks on HPC platforms''. M.A.M. would like to thank the organizers of the program and QWorld Association. 

\section*{Code and Data Availability}

{\raggedright
The sampling script, the sampled subset, the model and the training script are available at \href{\repo}{\texttt{https://github.com/saifibnaezhararko/}}\allowbreak\href{\repo}{\texttt{Accelerating-hybrid-quantum-classical-}}\allowbreak\href{\repo}{\texttt{machine-learning-tasks-on-HPC-platforms}}. 

\bibliographystyle{ieeetr}
\bibliography{references}

\end{document}